\documentclass[aps,twocolumn,floats,showpacs,superscriptaddress,floatfix]{revtex4}
\usepackage{epsfig}
\begin{document}

\title{Dislocation jamming and Andrade creep}

\author{M.-Carmen Miguel}
\affiliation{Departament de F\'{\i}sica Fonamental,
Facultat de F\'{\i}sica, Universitat de Barcelona, Av. Diagonal 647,
E-08028, Barcelona, Spain}
\author{Alessandro Vespignani}
\affiliation{The Abdus Salam International Centre
for Theoretical Physics, P.O. Box 586, 34100 Trieste, Italy}
\author{Michael Zaiser}
\affiliation{Center for Materials Science and Engineering, 
University of Edinburgh,
King's Buildings, Sanderson Building, Edinburgh EH93JL, UK}
\author{Stefano Zapperi}
\affiliation{INFM UdR Roma 1 and SMC, Dipartimento di Fisica, 
Universit\`a "La Sapienza", P.le A. Moro 2, 00185 Roma, Italy}
\date{\today}

\begin{abstract}

We simulate the glide motion of an assembly of interacting
dislocations under the action of an external shear stress and show
that the associated plastic creep relaxation follows Andrade's
law. Our results indicate that Andrade creep in plastically deforming
crystals involves the correlated motion of dislocation structures near
a dynamic transition separating a flowing from a jammed
phase. Simulations in presence of dislocation multiplication and noise
confirm the robustness of this finding and highlight the importance of
metastable structure formation for the relaxation process.

\end{abstract}

\pacs{61.72.Lk,81.40.Lm,68.35.Rh}

\maketitle

Andrade reported in 1910 that the creep deformation of soft metals at
constant temperature and stress grows in time according to a power law
with exponent $1/3$, i.e. $\gamma\sim t^{1/3}$ where $\gamma$ is the
global strain of the material \cite{Andrade10}. More generally, the
creep deformation curve usually follows the relation $\gamma=\gamma_0+
\beta t^{1/3}+\kappa t$, where $\gamma_0$ is the instantaneous plastic
strain, $\beta t^{1/3}$ is Andrade creep, and $\kappa t$ is referred
to as linear creep~\cite{Friedel,Cottrell}. This same behavior has
since been observed in many materials with rather different structures
leading to the conclusion that this should be a process determined by
quite general principles, independent of most material specific
properties. In crystalline materials, Andrade's and linear creep find
their microscopic origin in the dynamics of dislocations, the
topological defects responsible for their plastic
deformation~\cite{Friedel,Cottrell,Hirth92}. Plastic flow
only occurs when the externally applied stress overcomes a threshold
value, the yield stress of the material, such that large-scale
dislocation motion may take place. Despite various arguments proposed
within the dislocation
literature~\cite{Friedel,Cottrell,Mott53,Cottrell96,Nabarro97}, there
is no general consensus on the basic mechanism to explain Andrade's
law. Mott~\cite{Mott53} attributed the power law to an athermal
cooperative process taking place close to the yield stress, but his
idea was not worked out. Later explanations have always focused on
thermally activated processes~\cite{Cottrell96,Nabarro97} and rely on
assumptions that may not be fully warranted~\cite{disdep}.

Under the action of external stress, dislocations tend to glide
cooperatively due to their mutual long-range elastic interactions,
which can be attractive or repulsive. The anisotropic character of
dislocation-dislocation interactions contributes at the same time to
the formation of spatial dislocation structures observed in
transmission electron micrographs of plastically deformed
metals~\cite{Hahner98}. As a result of these peculiar interactions,
self-induced constraints build up in the system and the motion of
dislocations may eventually cease. Small variations of the external
loading, the density, or the dislocation distribution can, however,
enhance dislocation motion in a discontinuous manner. This gives rise
to rather complex and heterogeneous spatio-temporal patterns of
plastic flow, which have been observed experimentally in the form of
slip lines and slip bands emerging on the surface of metals
\cite{Becker32,Bengus67,NEU-83} or in the acoustic emission activity
of ice crystals~\cite{Miguel01}.  Other processes commonly taking
place in plastically deforming crystals such as hardening, fatigue, or
plastic instabilities \cite{ANA-99,DAN-00}, are further consequences
of the dislocations intriguing behavior.
 
In this Letter, we study the temporal relaxation of a relatively
simple dislocation dynamics model through numerical simulation. In
particular, we consider the behavior of parallel straight edge
dislocations moving in a single slip system under the action of
constant stress. We show that the model gives rise to Andrade-like
creep at short and intermediate times for a wide range of applied
stresses, without invoking thermally activated processes. At larger
times the strain-rate, which is proportional to the density of mobile
dislocations, crosses over to a linear creep regime (steady rate of
deformation), whenever the applied stress is larger than a critical
threshold $\sigma_c$, or decays exponentially to zero. These results
suggest that a possible interpretation of the creep laws could be
found within the general scenario wrapping a dynamic phase transition
from a flowing to a jammed dislocation phase. The ``jamming'' scenario
has been recently proposed~\cite{Liu} to understand a broad class of
non-equilibrium physical systems (granular media, colloids,
supercooled liquids, foams) which, in spite of their differences,
exhibit common properties such as slow dynamics and scaling features
near the jamming threshold. When jammed, these systems are unable to
explore phase space, but they can be unjammed by changing the stress,
the density, or the temperature. To further explore the analogies of
dislocation motion and these so-called jammed systems, we consider the
influence of dislocation multiplication, due for instance to the
activation of Frank-Read
sources~\cite{Friedel,Cottrell,Hirth92} during the
deformation process, and thermal-like fluctuations on the dynamics of
the dislocations. Dislocation multiplication favors the rearrangements
of the system and induces a linear creep regime (flowing phase) at
lower stress values, but it does not affect the initial power-law
creep. The introduction of a finite effective temperature $T$ has a
similar effect.

We consider a two-dimensional (2d) model representing a cross section
of a single-slip oriented crystal where $N$ point-like edge
dislocations glide in the $xy$ plane along directions parallel to the
$x$ axis. Dislocations with positive and negative Burgers vectors (the
topological charge characterizing a dislocation) ${\bf b}_n = \pm
b\hat{x}$ are assumed to be present in equal numbers, and the initial
number of dislocations is the same in every realization. Several
$2d$-models containing similar basic ingredients have been proposed in
the literature in the last few
years~\cite{Kubin87,Amodeo90,Groma93,Miguel01}.  An
important feature common to these models is that dislocations interact
with each other through the long-range elastic stress field they
produce in the host material. An edge dislocation with Burgers vector
$b \hat{x}$ located at the origin gives rise to a shear stress
$\sigma^{s}$ at a point ${\bf r}=(x,y)$ of the form
\cite{Friedel,Cottrell,Hirth92}
\begin{equation}\label{eq:1}
\sigma^{s}({\bf r})= D\frac{x(x^2-y^2)}{(x^2+y^2)^2},
\end{equation} 
where $D=\mu/2\pi(1-\nu)$ is a coefficient involving the shear modulus
$\mu$ and the Poisson ratio $\nu$ of the material.  We further assume
an overdamped dynamics in which the dislocation velocities are
linearly proportional to the local forces. Accordingly, the velocity
of the $n$th dislocation along the glide direction, if an external
shear stress $\sigma$ is also applied, is given by
\begin{equation}\label{eq:2}
\frac{\chi^{-1}_{\rm d}v_n}{b} = b_n (\sum_{m\neq n} \sigma^{s}({\bf r}_{nm}) + \sigma),
\end{equation}
where $\chi_{\rm d}$ is the effective mobility of the
dislocations~\cite{note} ($\chi^{-1}_{\rm d}/b$ the effective friction
per unit length) and ${\bf r}_{nm}\equiv\vec{r}_n-\vec{r}_m$ the
relative position vector of dislocations $n$ and $m$. Periodic
boundary conditions are imposed in the direction of motion (i.e. the
$x$ axis).  In order to take correctly into account the long range
nature of the elastic interactions (Eq.~\ref{eq:1}), we sum the stress
over an infinite number of images.  This sum can be performed exactly
and the results are reported in Ref.~\cite{Hirth92}.  When the
distance between two dislocations is of the order of a few Burgers
vectors, linear elasticity theory (i.e. Eq~.(\ref{eq:1})) breaks down.
In these instances, phenomenological nonlinear reactions, such as the
annihilation of a pair of dislocations, describe more accurately the
real behavior of dislocations in a crystal
\cite{Kubin87,Amodeo90,Groma93,Miguel01}. In our model, we
{\em annihilate} a pair of dislocations with opposite Burgers vectors
when the distance between them is shorter than $y_{\rm e}$ (for Cu,
$y_{\rm e} \approx 1.6$nm \cite{Essma79}). In the following, we
measure all lengths in units of $y_{\rm e}$, time in units of
$t_o=y_{\rm e}^2/(\chi D b^{3})$, and stresses in units of $Db/y_{\rm
e}$.

To analyze creep relaxation, we integrate numerically the $N$ coupled
equations using an adaptive step size fifth-order Runge-Kutta
algorithm. Simulations start from a configuration of $N_0$
dislocations randomly placed on a square cell of size $L$. We have
considered two different box sizes $L=100 y_{\rm e}$, and $L=300
y_{\rm e}$, with initial numbers of dislocations $N_0=400$, and
$N_0=1500$, respectively. We first relax the system until it reaches a
metastable arrangement, where the remaining dislocation density is
around $1$\%. Next we apply an external shear stress and let the
system evolve. The results are typically averaged over several
(100-400) random initial configurations.

\begin{figure}[t]
\centerline{\epsfig{file=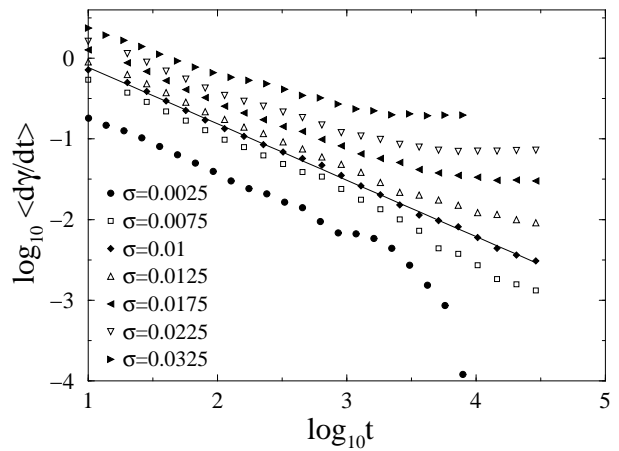,width=8cm,clip=!}}
\caption{The strain rate relaxation for different applied stresses at
$T=0$ for a system of size $L=100y_{\rm e}$. The initial density of
dislocations is around $1\%$. The solid line is the best linear fit of
the $\sigma=0.01$ curve and yields $d\gamma/dt\sim t^{-0.69}$.}
\label{fig1}
\end{figure}

\begin{figure}[t]
\centerline{\psfig{file=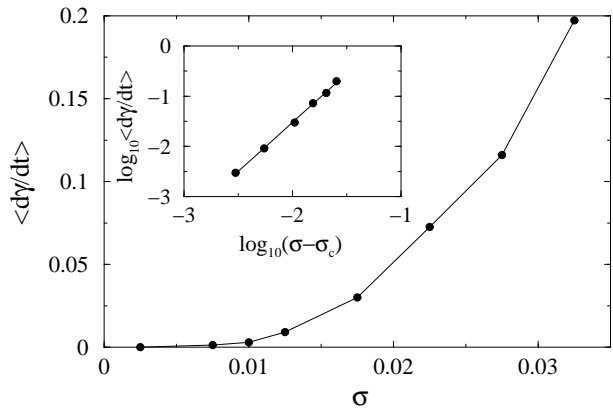,width=8cm,clip=!}}
\caption{The steady-state value of the strain rate as a function
of the applied stress. The inset shows that the strain rate
scales as $(\sigma-\sigma_c)^\beta$, with $\beta=1.8$.}
\label{fig2}
\end{figure}

In Fig.~\ref{fig1} we report the plastic strain rate of the material,
defined as $d\gamma/dt \sim \sum_i b_i v_i$, for different values of
the applied stress. The strain rate decays as a power law, with an
exponent close to $2/3$, in agreement with Andrade's law. For high
stresses the power law relaxation is followed by a plateau indicating
the onset of a linear creep regime. The crossover time increases as
the stress decreases, and for stresses lower than $\sigma \simeq
0.0075-0.01$ the plateau disappears and the creep decays exponentially
to zero. In Fig.~\ref{fig2} we display the steady-state strain rate as
a function of stress.  This plot suggests the presence of a
non-equilibrium phase transition between a moving and a jammed
stationary state controlled by the applied stress. Using $\sigma_c
=0.0075$, we find $d\gamma/dt \sim (\sigma-\sigma_c)^\beta$ with
$\beta= 1.8\pm 0.1$.

From our model and the analysis of our data, we propose a new
explanation of Andrade creep. In the course of time, most dislocations
tend to arrange themselves into metastable structures. These
structures often consist of small-angle dislocation boundaries
separating slightly misoriented crystalline blocks, whose stress
fields are screened on large length-scales, of dislocation dipoles, or
of far more complex dislocation arrangements~\cite{Miguel01}. The
stress field generated by a fraction of the dislocations in these
metastable configurations conserves, however, the initial long-range
character, and this enforces the continuation of the relaxation
process. If the applied stress is close to the critical threshold,
dislocations rearrange in a correlated and intermittent manner~\cite{Miguel01} 
exploring further configurations, yielding Andrade law as the average
result. Above the critical stress, 
the system eventually enters a linear creep regime in which
dislocation structures are coherently moving with a velocity that
grows with the applied stress. On the contrary, for stress values
below the critical threshold, dislocations cannot get around the
self-induced constraints built up by their mutual
(attractive/repulsive) interactions and thus they are unable to
further explore configuration space, the signature of a jammed
system~\cite{Liu}. The closer is the applied stress to the threshold,
the longer is the extent of the power-law regime before the system
falls either in the flowing or in the jammed state. Precisely at the
critical point and in an infinite system, the Andrade power-law would
last indefinitely.

Within the proposed scenario, it appears natural to ask whether two
important effects, namely dislocation multiplication and thermal
fluctuations, affect this behavior in a relevant way. During plastic
deformation, in fact, new dislocations are created within the
crystal. It is widely believed that the Frank-Read multiplication
mechanism~\cite{Friedel,Cottrell,Hirth92} is the most
relevant for dislocations gliding under creep deformation. Since
Frank-Read sources cannot be directly simulated in a two dimensional
model, we employ a phenomenological procedure, introducing
dislocations pairs with a rate $r$ proportional to the external
stress.  Similar multiplication mechanisms have been successfully used
in the past
\cite{Kubin87,Amodeo90,Groma93,Miguel01}. Thermal
fluctuations are accounted for by adding a Gaussian random force per
unit length $\eta_n(t)/b$ to Eq.~\ref{eq:2}. The force has zero mean
and its correlations are given by
\begin{equation}
\langle \eta_n(t)\eta_m(t')\rangle = k_B T \chi_{\rm d}^{-1} \delta_{n,m}\delta(t-t'), 
\end{equation}
where $T$ is an effective temperature characterizing the strength of
the fluctuations \cite{note1}. 
This random force could also mimic, as a first
approximation, the influence of dislocation motion in other slip
systems that may be active simultaneously in the material
(see~\cite{Hahner98,Groma00}).

\begin{figure}
\centerline{\psfig{file=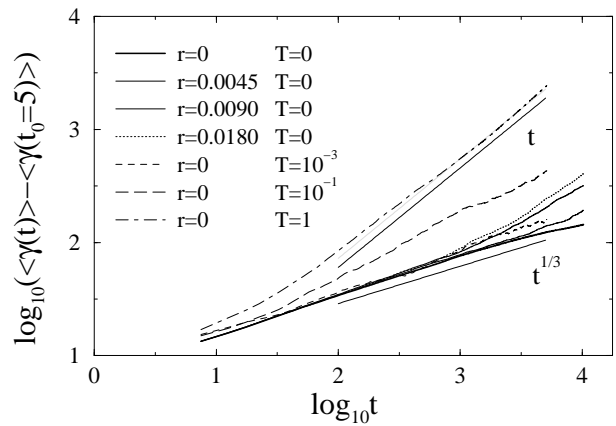,width=8cm,clip=!}}
\caption{The effect of the dislocation multiplication rate $r$ and of
the effective temperature $T$ on the plastic strain in a system of
size $L=100y_{\rm e}$ with an applied stress $\sigma=0.0075$.  Andrade
creep ($\gamma\sim t^{1/3}$) crosses over to linear creep ($\gamma\sim
t$) at large times, with a crossover depending on $r$ and $T$.}
\label{fig3}
\end{figure}

\begin{figure}[t]
\centerline{\psfig{file=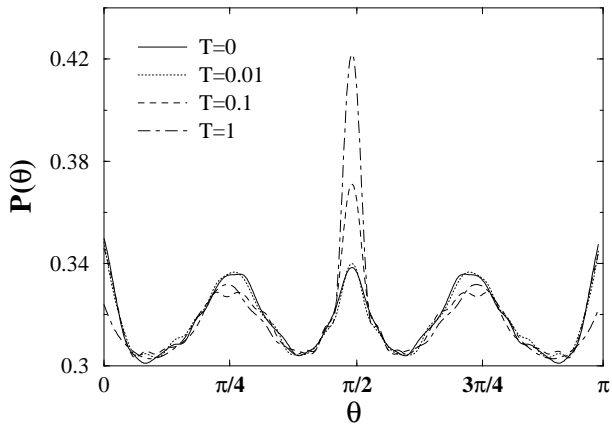,width=8cm,clip=!}}
\caption{The angle distribution as a function of the temperature at
$\sigma=0.0075$.}
\label{fig4}
\end{figure}

We have performed a series of simulations for different multiplication
rates and effective temperatures (measured in units of $D b^3/k_B$)
when the applied stress is close to the critical value
(i.e. $\sigma=0.0075$). We used the same procedure and initial
conditions described previously and, given the additional fluctuations
present in the strain rate, we focus our attention on the integrated
plastic strain $\gamma$. As shown in Fig.~\ref{fig3}, a linear creep
regime appears after a crossover time which decreases with $r$ and
$T$. Nevertheless Andrade creep (i.e. $\gamma\sim t^{1/3}$) still
persists at shorter times. A visual inspection of the dislocation
arrangements during the deformation is useful to understand the origin
of the various creep regimes. At low $T$, we observe the presence of
slowly relaxing metastable structures as for $T=0$.  On the other
hand, at much higher effective temperature (i.e. $T \simeq 1$) these
structures have completely broken up (melt), the flow of dislocations
becomes fluid-like, and Andrade creep disappears. In
order to substantiate quantitatively this statement, we measure the
angular correlation between dislocations. For each dislocation pair
with coordinates $\vec{r}_i$ and $\vec{r}_j$, we define $\theta$ as
the azimuthal angle with respect to the $y$ axis of the vector
$\vec{r}_{ij}$ (i.e.  $\theta\equiv\arccos(\hat{y}\cdot
\vec{r}_{ij}/|\vec{r}_{ij}|)$).  Thus two dislocations placed in the
same wall have either $\theta\simeq 0$ or $\theta\simeq \pi$, a
dislocation dipole is characterized by $\theta\simeq \pi/4, 3\pi/4$
and two dislocations in a pileup yield $\theta\simeq
\pi/2$~\cite{Friedel,Cottrell,Hirth92}.  The distribution
$P(\theta)$ is obtained after averaging over all dislocation pairs in
several realizations ($\sim 100$) of the dynamics at a given instant.
Fig.~\ref{fig4} shows $P(\theta)$ for different values of $T$.  At low
effective temperatures we observe five roughly equivalent peaks,
corresponding to walls, dipoles and pileup configurations. As the
effective temperature increases, the peaks at $\theta=0,\pi$ and
$\theta\simeq \pi/4, 3\pi/4$, corresponding to walls and dipoles,
progressively disappear, while the $\theta=\pi/2$ peak remains. This
peak indicates that even in the high temperature regime the
dislocations are correlated in the $x$ direction, 
the only possible direction of motion. Apart
from this no other structure remains in the system.

In conclusion, our results capture 
several features of the phenomenology observed
in creep deformation experiments.
Andrade's scaling and the creep curve appear to be
controlled by a non-equilibrium phase transition between a jammed
and a flowing dislocation state. 
Determining the  critical stress value,  and characterizing its
behavior (possible dependence on density and temperature) is of utmost
importance for practical purposes, since it establishes the mechanical
strength of crystalline materials.

We thank R. Pastor-Satorras for useful
discussions. M.C.M. acknowledges financial support from the Ministerio
de Ciencia y Tecnolog\'{\i}a (Spain).


\begin{thebibliography}{47}
\bibitem{note} The effective mobility of the dislocations may be
controlled by thermal activation, giving rise to a
temperature-dependent mobility $\chi_{\rm d}$. This, however, 
does not change the nature of the $T=0$ non-equilibrium critical point.
\bibitem{note1} We consider a model of point dislocations,
(i.e rigid lines that thread the whole
sample). Thus the thermal effects we have included may only be relevant
in very thin samples, thicker specimens being captured by 
zero temperature results.
\end{thebibliography}
\end{document}